\documentclass[aps,prc,floatfix,showpacs,preprintnumbers,amsmath,amssymb,superscriptaddress,reprint]{revtex4-1}

\usepackage[bookmarks=false]{hyperref}
\usepackage{latexsym}
\usepackage{verbatim}
\usepackage{graphics}
\usepackage{subcaption}
\usepackage{amsmath}
\usepackage{setspace}
\usepackage{graphicx,amsmath,color}
\usepackage{todonotes}
\usepackage{placeins}
\usepackage{afterpage}
\usepackage[labelsep=period]{caption}

\makeatletter
\g@addto@macro\bfseries{\boldmath}
\makeatother

\bibpunct{[}{]}{,}{n}{}{} 
\usepackage[resetlabels]{multibib}

\definecolor{lightred}{rgb}{1,0.5,0.5}
\definecolor{lightgreen}{rgb}{0.5,1,0.5}
\definecolor{darkgreen}{rgb}{0.5,0.9,0.5}

\definecolor{Mycolor2}{HTML}{0c9008}

\begin{document}

\title{Measurement of the Photon Beam Asymmetry \\in $\vec{\gamma} p\to K^+\Sigma^0$ at $E_{\gamma} = 8.5$\,GeV}

\affiliation{Arizona State University, Tempe, Arizona 85287, USA}
\affiliation{National and Kapodistrian University of Athens, 15771 Athens, Greece}
\affiliation{Carnegie Mellon University, Pittsburgh, Pennsylvania 15213, USA}
\affiliation{The Catholic University of America, Washington, D.C. 20064, USA}
\affiliation{University of Connecticut, Storrs, Connecticut 06269, USA}
\affiliation{Duke University, Durham, North Carolina 27708, USA}
\affiliation{Florida International University, Miami, Florida 33199, USA}
\affiliation{Florida State University, Tallahassee, Florida 32306, USA}
\affiliation{The George Washington University, Washington, D.C. 20052, USA}
\affiliation{University of Glasgow, Glasgow G12 8QQ, United Kingdom}
\affiliation{GSI Helmholtzzentrum f\"ur Schwerionenforschung GmbH, D-64291 Darmstadt, Germany}
\affiliation{Institute of High Energy Physics, Beijing 100049, People's Republic of China}
\affiliation{Indiana University, Bloomington, Indiana 47405, USA}
\affiliation{Alikhanov Institute for Theoretical and Experimental Physics NRC Kurchatov Institute, Moscow 117218, Russia}
\affiliation{Thomas Jefferson National Accelerator Facility, Newport News, Virginia 23606, USA}
\affiliation{Forschungszentrum Juelich Nuclear Physics Institute, 52425 Juelich, Germany}
\affiliation{University of Massachusetts, Amherst, Massachusetts 01003, USA}
\affiliation{Massachusetts Institute of Technology, Cambridge, Massachusetts 02139, USA}
\affiliation{National Research Nuclear University Moscow Engineering Physics Institute, Moscow 115409, Russia}
\affiliation{Norfolk State University, Norfolk, Virginia 23504, USA}
\affiliation{North Carolina A\&T State University, Greensboro, North Carolina 27411, USA}
\affiliation{University of North Carolina at Wilmington, Wilmington, North Carolina 28403, USA}
\affiliation{Northwestern University, Evanston, Illinois 60208, USA}
\affiliation{Old Dominion University, Norfolk, Virginia 23529, USA}
\affiliation{University of Regina, Regina, Saskatchewan, Canada S4S 0A2}
\affiliation{Universidad T\'ecnica Federico Santa Mar\'ia, Casilla 110-V Valpara\'iso, Chile}
\affiliation{Tomsk State University, 634050 Tomsk, Russia}
\affiliation{Tomsk Polytechnic University, 634050 Tomsk, Russia}
\affiliation{A. I. Alikhanian National Science Laboratory (Yerevan Physics Institute), 0036 Yerevan, Armenia}
\affiliation{College of William and Mary, Williamsburg, Virginia 23185, USA}
\affiliation{Wuhan University, Wuhan, Hubei 430072, People's Republic of China}
\author{S.~Adhikari}
\affiliation{Old Dominion University, Norfolk, Virginia 23529, USA}
\author{A.~Ali}
\affiliation{GSI Helmholtzzentrum f\"ur Schwerionenforschung GmbH, D-64291 Darmstadt, Germany}
\author{M.~Amaryan}
\email[Corresponding author:]{mamaryan@odu.edu}
\affiliation{Old Dominion University, Norfolk, Virginia 23529, USA}
\author{A.~Austregesilo}
\affiliation{Carnegie Mellon University, Pittsburgh, Pennsylvania 15213, USA}
\author{F.~Barbosa}
\affiliation{Thomas Jefferson National Accelerator Facility, Newport News, Virginia 23606, USA}
\author{J.~Barlow}
\author{E.~Barriga}
\affiliation{Florida State University, Tallahassee, Florida 32306, USA}
\author{R.~Barsotti}
\affiliation{Indiana University, Bloomington, Indiana 47405, USA}
\author{T.~D.~Beattie}
\affiliation{University of Regina, Regina, Saskatchewan, Canada S4S 0A2}
\author{V.~V.~Berdnikov}
\affiliation{National Research Nuclear University Moscow Engineering Physics Institute, Moscow 115409, Russia}
% ...
\collaboration{The \textsc{GlueX} Collaboration}

\vskip 1.50in
\vskip 2in
\date{\today}

\begin{abstract}
\vskip 0.2in
We report measurements of the photon beam asymmetry $\Sigma$ for the reaction $\vec{\gamma} p\to K^+\Sigma^0$(1193) using the GlueX spectrometer in Hall D at Jefferson Lab. Data were collected using a linearly polarized photon beam in the energy range of 8.2-8.8\,GeV incident on a liquid hydrogen target.  The beam asymmetry $\Sigma$ was measured as a function of the Mandelstam variable $t$, and a single value of $\Sigma$ was extracted for events produced in the $u$-channel. These are the first exclusive measurements of the photon beam asymmetry $\Sigma$ for the reaction in this energy range. For the $t$-channel, the measured beam asymmetry is close to unity over the $t$-range studied, $-t=(0.1-1.4)~$(GeV/$c$)$^{2}$, with an average value of $\Sigma$~=~1.00~$\pm$~0.05.
This agrees with theoretical models that describe the reaction via the natural-parity exchange of the $K^{*}$(892) Regge trajectory. A value of $\Sigma$ = 0.41~$\pm~$0.09 is obtained for the $u$-channel  integrated up to  $-u=2.0$~(GeV/$c$)$^{2}$.

\end{abstract}

\maketitle

\section{Introduction}

The GlueX experiment at Thomas Jefferson National Accelerator Facility (Jefferson Lab) was designed to study the light quark meson spectrum and to search for exotic resonances. It uses a high-intensity linearly-polarized photon beam impinging on a liquid hydrogen target and is able to access a broad range of final states. The interpretation of experimental data from photoproduction of pseudoscalar mesons requires a deep understanding of the production mechanism, which is complicated by the possible excitation of baryon resonances. In this experiment, we study photoproduction of the strange pseudoscalar meson $K^+$ in the $\vec{\gamma} p\to K^+\Sigma^0$ reaction, above the baryon resonance region. While the high-energy domain in photoproduction of pseudoscalar mesons is relatively well understood in the framework of Regge theory, precise experimental data for the photoproduction of many different final states at high energy are scarce. In this analysis, we focus on the photoproduction reaction $\vec{\gamma} p \rightarrow K^+\Sigma^0$ to study the mechanism of strange Reggeon exchange and measure the relative contributions of natural and unnatural parity exchange via beam asymmetry measurements.

Our understanding of the photoproduction of kaons at these energies is based predominantly on measurements from SLAC~\cite{Quinn1975,Quinn}. These measurements were not fully exclusive - the beam was untagged bremsstrahlung and only the final state $K^+$ was detected. The first paper reported measurements of beam asymmetry for the sum of the two photoproduction reactions, $K^+\Lambda$ and $K^+\Sigma$. It was found to be close to unity. In the later paper, the authors used the ratio of the cross sections, which was also close to unity, to extract separate asymmetries for the two processes as a function of $t$-Mandelstam. Prior to this current publication, these were the only available measurements above the baryon resonance region.
 
Theoretical models~\cite{Gold,Levy, Guidal,Ghent,Ghent2} are necessary for extracting 
information from the more detailed measurements obtained at lower beam energy, such as the beam asymmetry measurements from both proton and neutron targets with a photon beam at 1.5-2.4\,GeV by LEPS~\cite{Zegers:2003ux, Kohri:2006yx}, the measurements below 1.5\,GeV at GRAAL~\cite{Lleres:2007tx,Lleres:2008em}, and the recent CLAS results~\cite{ Paterson:2016vmc}, which provide extensive measurements of many observables for hadronic mass $W$ from 1.71 to 2.19\,GeV.  

These measurements have been important for resolving new states and also the status of many excited baryon states, however the precision of the existing high-energy data limited the accuracy of some of the modeling needed for the baryon studies.  The new and more precise data reported here will make an impact on models used in the lower energy studies. 

Below, we present the first exclusive measurement of the photon beam asymmetry $\Sigma$ in the reaction $\vec{\gamma} p \to K^+\Sigma^0$ beyond the resonance region. The analysis was performed with approximately 20\% of the data collected in the first phase of the GlueX experiment, which corresponds to a luminosity of 20.8~pb$^{-1}$ in the beam energy range between 8.2 and 8.8\,GeV.

\section{Theory}
The Mandelstam variables $s$, $t$ and $u$ in the reaction $\vec{\gamma} p\to K^{+}\Sigma^{0}$ are defined as:
\begin{align}
s&=(p_\mathrm{beam}+p_\mathrm{target})^{2},\\
t&=(p_\mathrm{beam}-p_{K^+})^{2},\\
u&=(p_\mathrm{target}-p_{K^{+}})^{2},
\end{align}
where $p_\mathrm{beam}$,\,$p_\mathrm{target}$ and $p_{K^{+}}$ are the four-momenta of the incoming photon beam, the target proton and the produced $K^+$ meson respectively.

The observables of the photoproduction reaction are discussed in terms of $s$-channel helicity amplitudes with definite parity in the $t$-channel to leading order in  $s$ defined in Ref.~\cite{Gold}: 
\begin{eqnarray}
f_1 = f_{1+,0+},\nonumber\\
f_2 = f_{1+,0-},\nonumber\\
f_3 = f_{1-,0+},\nonumber\\
f_4 = f_{1-,0-},
\end{eqnarray}
where in $f_{ab,cd}$ the subscripts $a,b,c,d$ 
represent the helicities of the incoming photon, the target proton, the produced spin-zero meson and the recoiling baryon, respectively. 
The following combinations can be formed:
\begin{eqnarray}
f_1^{\pm} = \frac{1}{2}(f_1\pm f_4),\nonumber\\
f_2^{\pm}=\frac{1}{2}(f_2\mp f_3),
\end{eqnarray}
where the superscript $+(-)$ indicates natural (unnatural) parity exchange in the $t$-channel. In Regge theory, for the reaction of interest, $\vec{\gamma} p \to K^+\Sigma^0$, these are realized via exchange of $K^{*}(892)$ and $K(494)$  trajectories for the natural and unnatural parity exchanges, respectively.

The polarized photon beam asymmetry is given by 
\begin{align}
\Sigma &= \Big[\frac{d\sigma_{\perp}}{dt}-\frac{d\sigma_{\parallel}}{dt}\Big ]\Big /\Big[\frac{d\sigma_{\perp}}{dt}+\frac{d\sigma_{\parallel}}{dt}\Big ]\nonumber\\
&=\frac{(\vert f_1^+\vert^2+\vert f_2^+\vert^2-\vert f_1^-\vert^2 -\vert f_2^-\vert ^2)}{
(\vert f_1^+\vert^2+\vert f_2^+\vert^2+\vert f_1^-\vert^2 +\vert f_2^-\vert ^2)},
\label{eqn:parity}
\end{align}
where $\frac{d\sigma_{\perp}}{dt}$ ($\frac{d\sigma_{\parallel}}{dt}$) is the cross section with a photon beam polarized perpendicular (parallel) to the reaction plane. 
The experimental value of $\Sigma$ provides a direct measurement of the relative contributions of the natural and unnatural parity exchange mechanisms to the photoproduction of the $K^+\Sigma^0$ final state.  
\section{Experiment}

The measurements were performed using the GlueX spectrometer, which is located in Hall D at Jefferson Lab. An 11.6\,GeV electron beam from the Continuous Electron Beam Accelerator Facility is used to create a tagged linearly polarized photon beam by coherent bremsstrahlung off a diamond radiator. The polarization approaches 40\% in the region of the coherent peak, from 8.2 to 8.8\,GeV. The scattered electrons are directed into the Tagger Detector, a scintillating-fiber array which, by measuring the momenta of the recoil electrons, enables a measurement of the energy of the produced photons to 0.1\% precision within the region of the coherent peak.

The photon beam passes through a collimator in order to suppress the incoherent part, a triplet polarimeter~\cite{Dugger:2017zoq} and a pair spectrometer~\cite{Barbosa:2015bga}, which provide continuous, non-invasive measurements of the photon beam polarization and the relative flux, respectively, before reaching the liquid hydrogen target. 
The target is surrounded by a scintillator start counter~\cite{Pooser:2019rhu}, a straw-tube central drift chamber~\cite{CDC_nim} and a lead and scintillating-fiber barrel calorimeter~\cite{Beattie:2018xsk}, all inside the bore of a superconducting solenoid.
Four sets of planar wire drift chambers~\cite{Pentchev:2017omk} are also located inside the solenoid, downstream of the central drift chamber. 
A time-of-flight scintillator wall and a forward lead-glass calorimeter~\cite{Moriya:2013aja} are located further down the beamline and outside of the solenoid.  
The drift chambers provide measurements of momentum and specific energy loss for charged particles, while the calorimeters provide energy and position measurements 
for showers caused by both charged and neutral particles. Time-of-flight measurements for particle identification are provided by the start counter, the calorimeters and the time-of-flight wall. The trigger signal is generated for events that deposit sufficient energy in the calorimeters.
The spectrometer has a nearly hermetic angular coverage. 

The data used in this analysis were collected in spring 2017. Four orientations of the diamond radiator were used to produce bremsstrahlung photons in two sets with
orthogonal linear polarization, one set parallel and perpendicular to the lab floor (referred
to as ‘0/90’), and a second set, rotated by 45$^{\circ}$
from the first one (‘-45/45’). The two different sets of orientations allow an independent check of systematic uncertainties.

\section{Event Selection}

The exclusive reaction $\vec{\gamma} p \to K^+\Sigma^0$ was selected using the subsequent decays of $\Sigma^0\to\Lambda^0\gamma$ and $\Lambda^0\to p\pi^-$. Candidate events for this reaction were required to contain at least two positively charged tracks, one negative track and one photon candidate. Extra tracks, showers and tagged beam photons were also permitted in the initial event selection. The proton was identified via its specific energy loss $dE/dx$ in the central drift chamber, and time-of-flight information was used to refine the selection of all the charged-particle tracks. The absolute value of the squared missing mass for the reaction was limited to less than 0.08~(GeV/$c^2$)$^2$. A kinematic fit was used to select particle combinations satisfying conservation of energy and momentum with a constraint on the event vertex. Following the kinematic fit, further event selection required that the vertex of the $K^+$ track originate within 1~cm from the beamline and within the target volume, while the pion and proton from the $\Lambda$ decay were permitted to originate from a detached vertex. A quality requirement was placed on neutral showers in the forward calorimeter in order to reduce the likelihood that they were caused by split-off clusters from charged particle showers~\cite{Barsotti:2020}.         

\begin{figure} [!ht]
\centering
\includegraphics[width=0.5\textwidth]{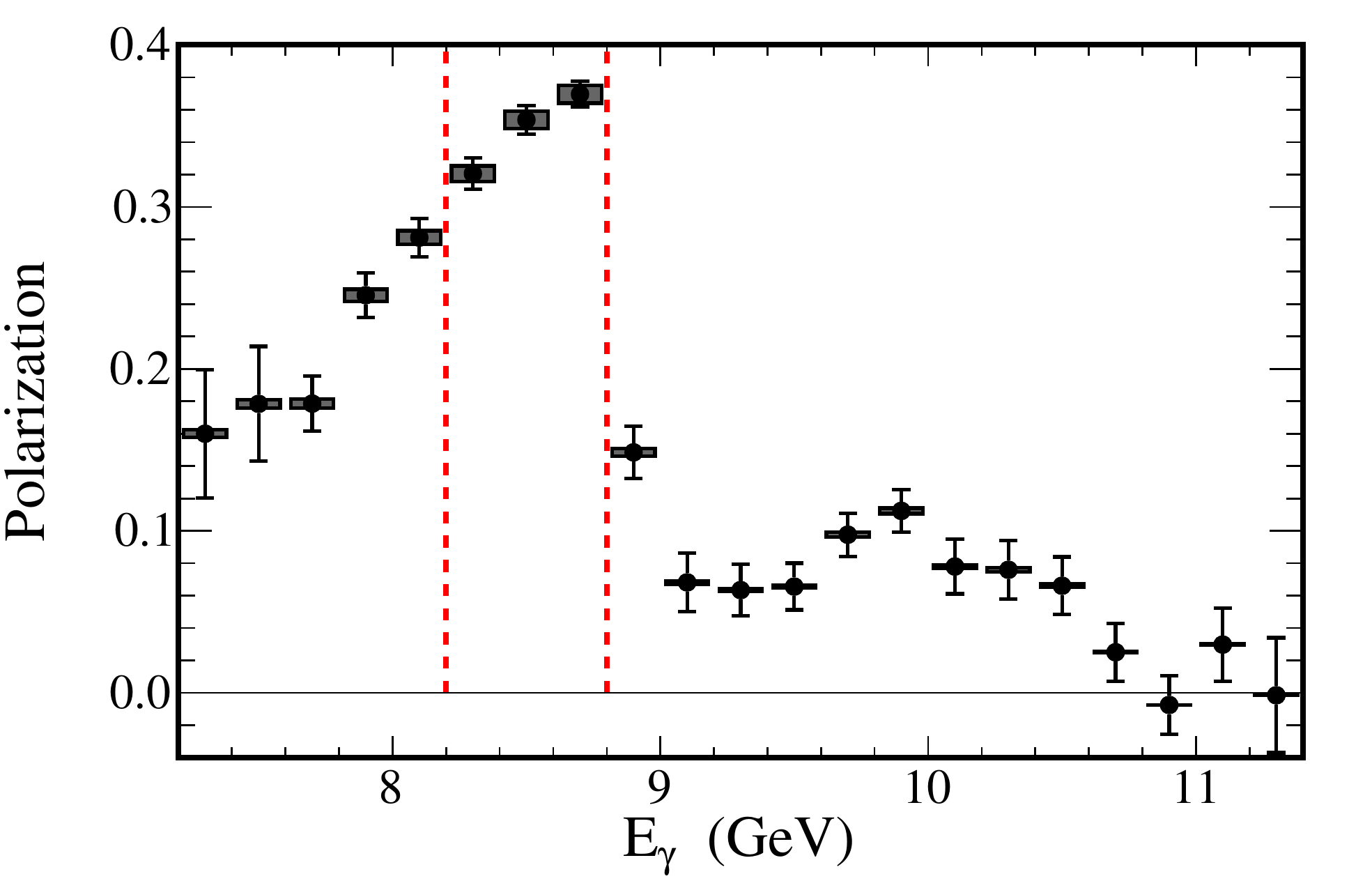}
\caption{Photon beam polarization as a function of beam energy, as measured by the triplet polarimeter, averaged over the four different diamond orientations.  Dashed vertical lines indicate the beam energy range used for this analysis.  Vertical error bars show the statistical uncertainty and inner shaded regions show the systematic uncertainty due to the 1.5\% relative uncertainty from the polarimeter analyzing power.  The polarizations for the individual orientations are presented in Ref.~\cite{Adhikari:2019gfa}.}
\label{fig:pol}
\end{figure}

The beam photons were selected from the coherent peak region, between 8.2 and 8.8\,GeV, where the polarization was highest. Figure~\ref{fig:pol} shows the measured polarization as a function of the photon energy averaged over the four diamond orientations. Dashed vertical lines indicate the photon beam energy range used in this analysis.

The energy of the beam photon initiating the event was defined by the position of the fired tagger counter in the Tagger Detector.  The candidates were selected using the time difference $|\Delta t|$ between the timing of the signal in the counter, projected to the vertex location, and the vertex time.  The electron beam had a 4\,ns bunch structure but the vertex timing resolution permitted the association of the events with a particular bunch, thus improving the $|\Delta t|$ resolution.  Prompt beam candidates were selected in the range $|\Delta t|<2$\,ns.  Accidental coincidences, mostly within the same bunch, would provide incorrect beam energies.  Such background was statistically subtracted by selecting a sample of out-of-time candidates in the window  $6\,\text{ns}<|\Delta t|<18\,\text{ns}$.

\begin{figure}[!ht]
\centering
\includegraphics[width=0.5\textwidth]{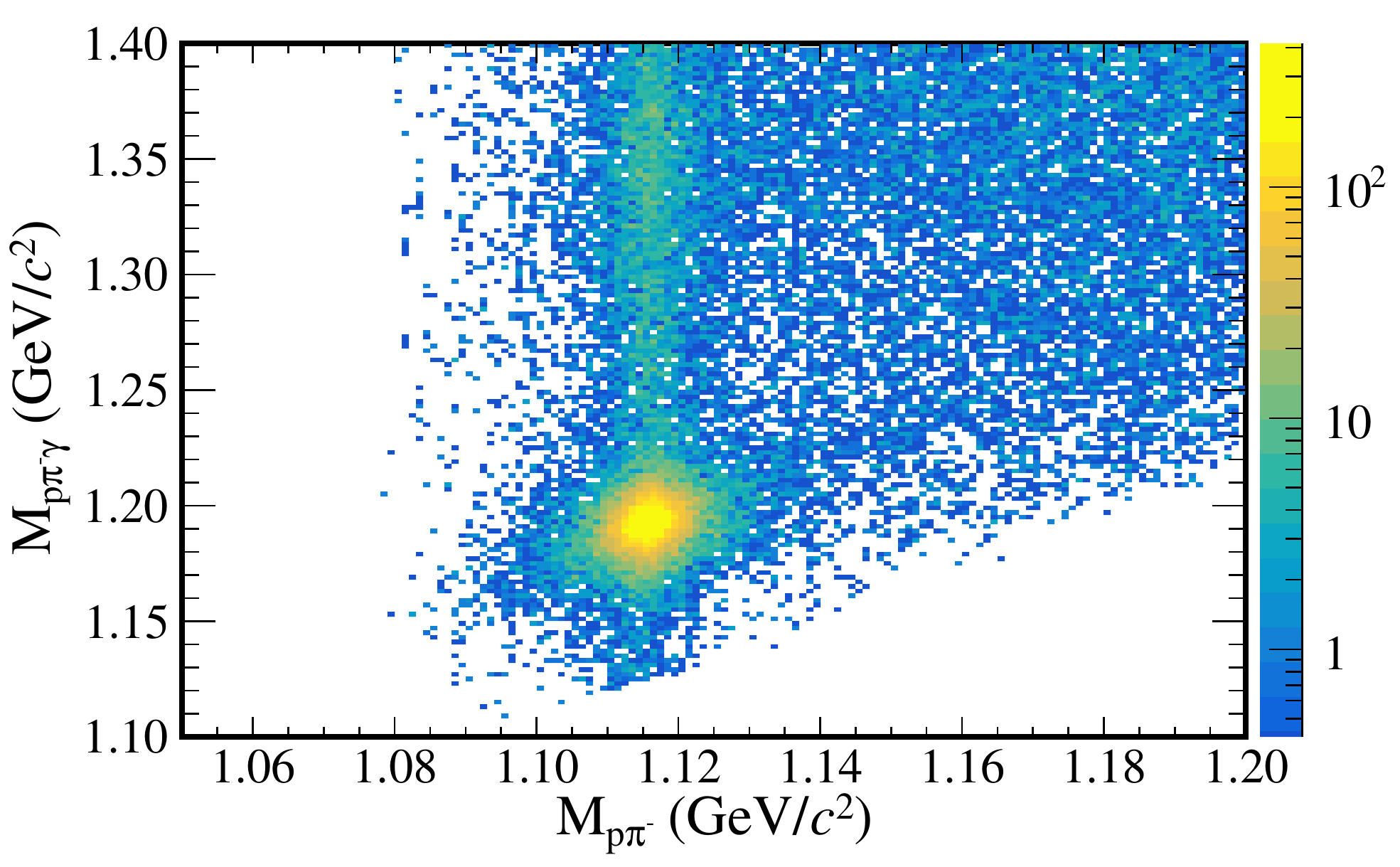}
\caption{ Invariant mass of $p\pi^{-}\gamma$ vs. invariant mass of $p\pi^{-}$ after all cuts. The enhancement in the overlap region corresponds to the $\Lambda^0(1116)$ and $\Sigma^{0}(1193)$.}
\label{fig:mass_2d}
\end{figure}

\begin{figure}[!ht]
\centering
\includegraphics[width=0.5\textwidth]{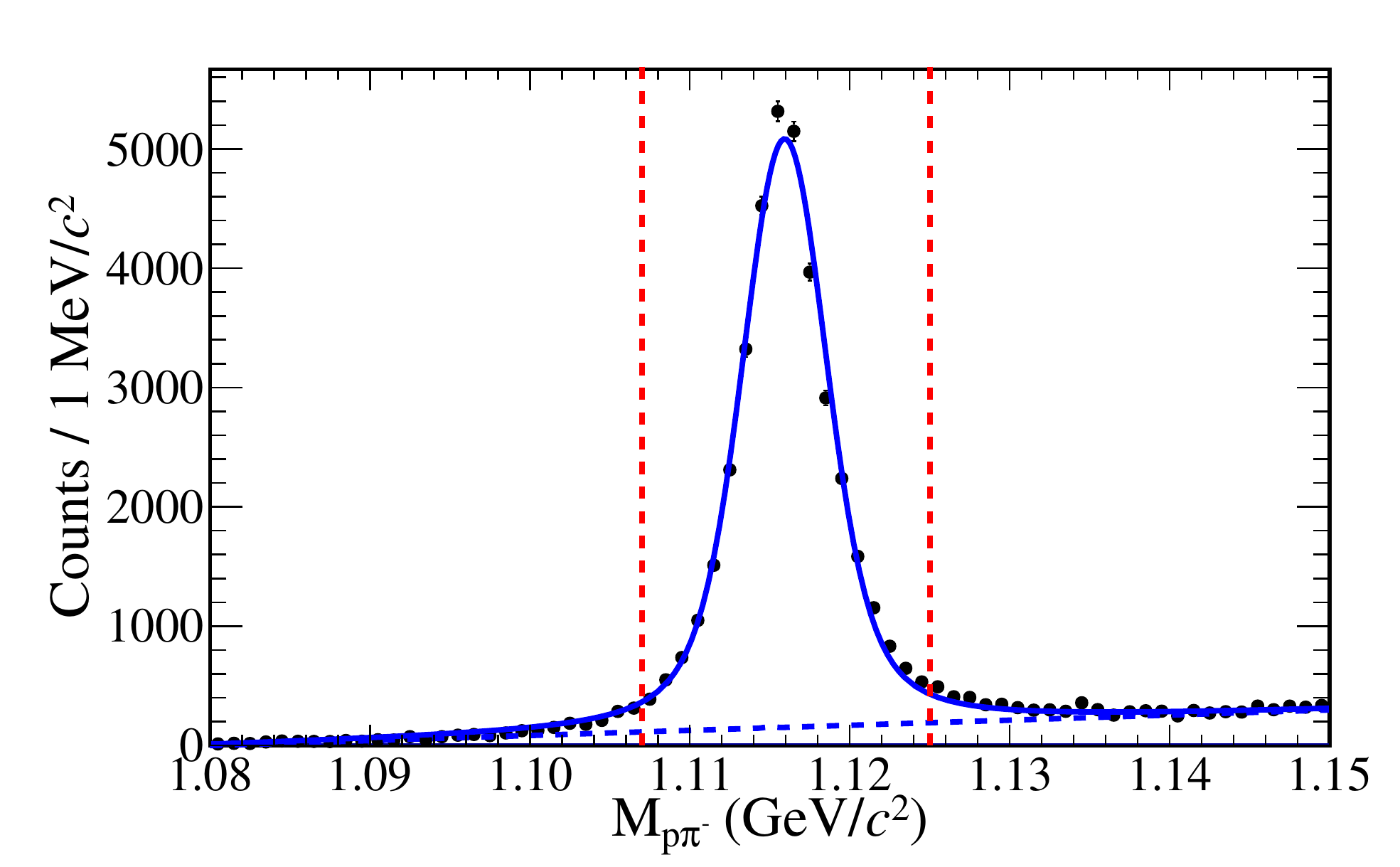}
\caption{Invariant mass of $p\pi^{-}$ (solid circles). The solid curve is the sum of a Voigtian and a first order Chebyshev polynomial (dashed curve) fitted to the data. The selection region of $\Lambda$ signal events is indicated by the vertical dashed lines.}
\label{fig:lambda}
\end{figure}

Figure~\ref{fig:mass_2d} shows the correlation between the invariant mass of the $p\pi^{-}\gamma$ system and its $p\pi^{-}$ subsystem. A clear enhancement can be seen in the  overlap  region between the masses of $\Sigma^{0}(1193)$ and $\Lambda^0 (1116)$, respectively. The one-dimensional $p\pi^-$ mass distribution in Fig.~\ref{fig:lambda} shows the $\Lambda$ peak. This distribution was fitted using a Voigtian function for the signal and a first order Chebyshev polynomial for the background. The fit shows a mean value of $M_{\Lambda}$~=~1116 MeV/$c^2$ and has a corresponding Gaussian width $\sigma$~=~3\,MeV/$c^2$. 
Events within the range $\vert{M_{p\pi^{-}} - M_{\Lambda}\vert} < 3\sigma$  were selected as shown in Fig.~\ref{fig:lambda} by dashed vertical lines to reconstruct the invariant mass of $\Lambda\gamma$.

\begin{figure}
\centering
\includegraphics[width=0.5\textwidth]{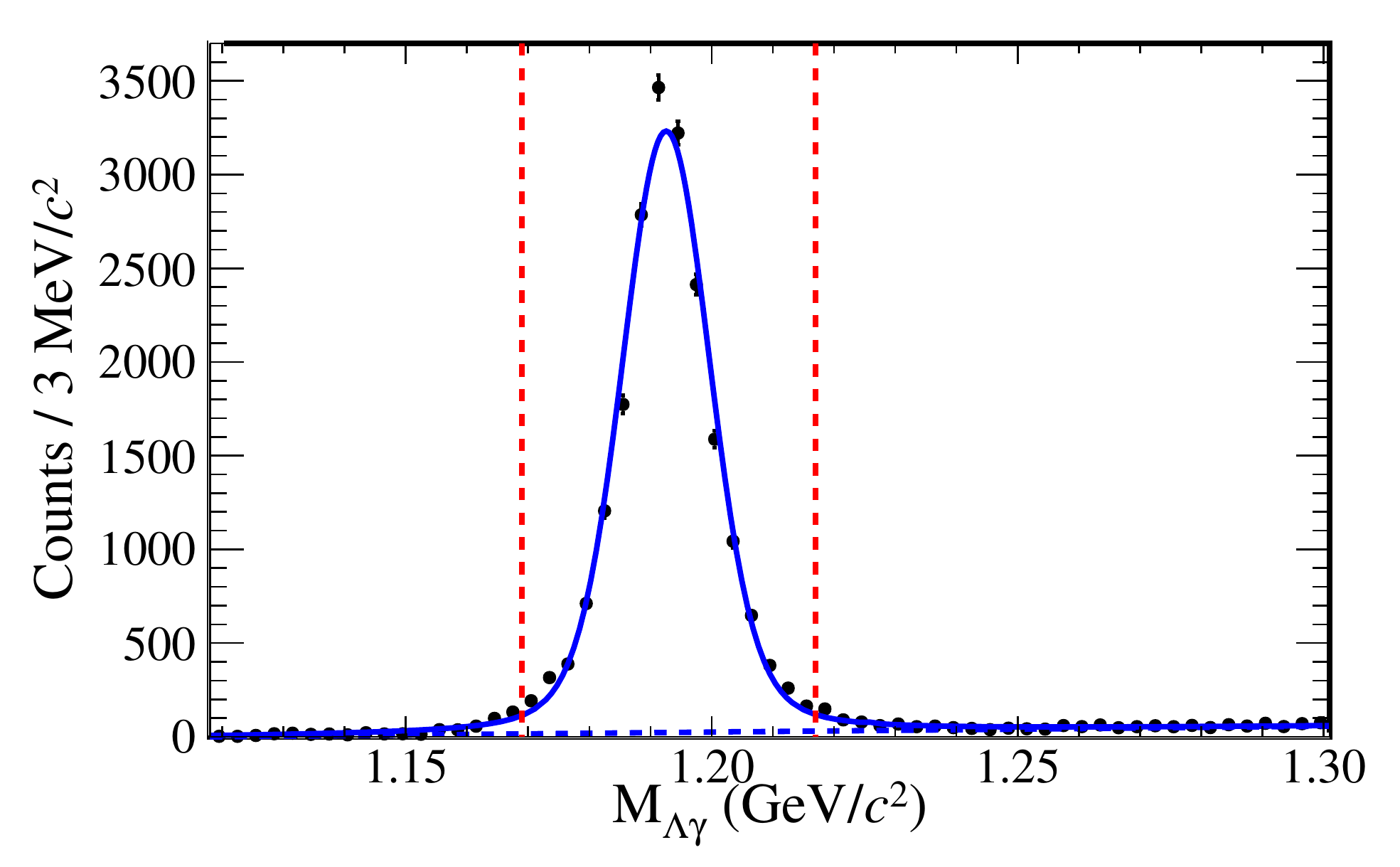}
\caption{~The invariant mass of $\Lambda\gamma$ (solid circles).  The solid curve is the sum of a Voigtian and a first order Chebyshev polynomial (dashed curve) fitted to the data. The selection region of $\Sigma^0$ events for further analysis is indicated by the vertical dashed lines.}
\label{fig:sigma_fit}
\end{figure}

Figure~\ref{fig:sigma_fit} shows the invariant mass of the $\Lambda\gamma$ system. 
It was fitted using a Voigtian function for the signal and a first order Chebyshev polynomial for the background. 
A mean value of $M_{\Sigma}$~=~1193 MeV/$c^2$ and a corresponding Gaussian width of $\sigma$~=~8\,MeV/$c^2$ were obtained for the $\Sigma^{0}$ peak.  
The range $\vert{M_{\Lambda\gamma} - M_{\Sigma}\vert} < 3\sigma$~is indicated by dashed vertical lines. 
The events within this range, 1.169\,GeV/$c^2 < M_{\Lambda\gamma}<$ 1.217\,GeV/$c^2$, were used for the beam asymmetry analysis. The fraction of background events within 3$\sigma$ of the peak was found to be approximately constant with $t$ at about 2\%.  

\begin{figure}[htb]
\includegraphics[width=0.45\textwidth]{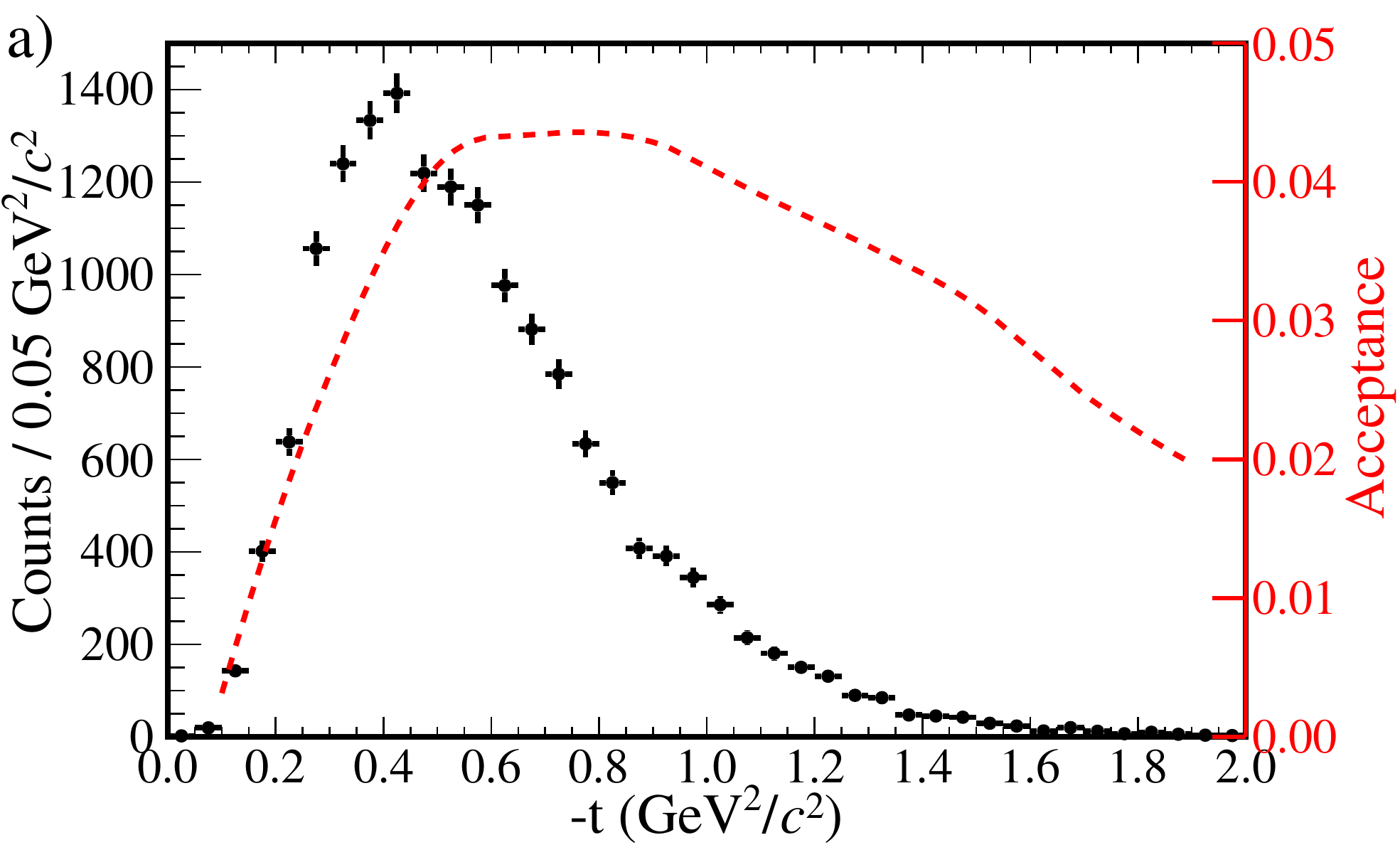}
\includegraphics[width=0.45\textwidth]{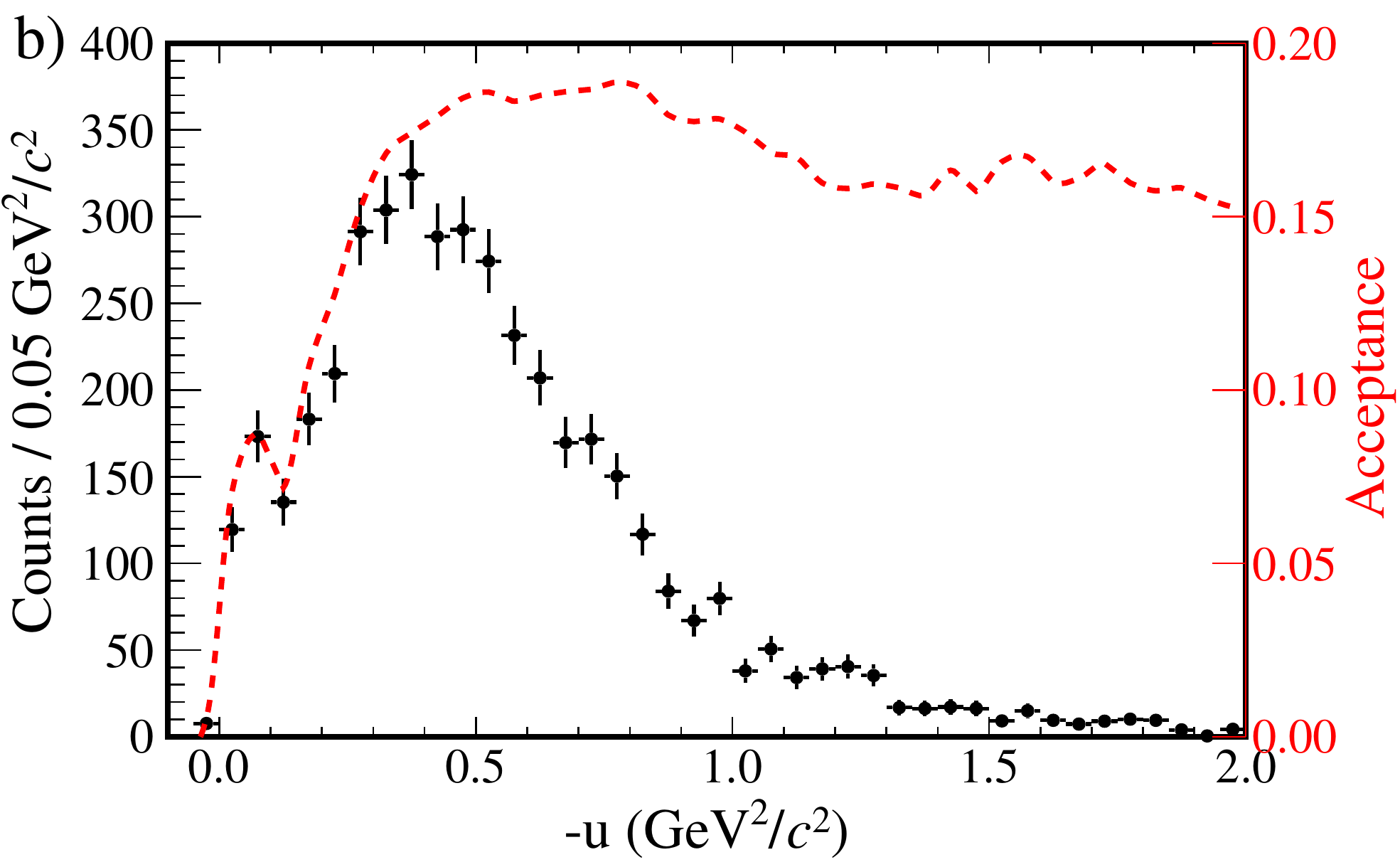}
\caption{Event yields for $\vec{\gamma} p\to K^+\Sigma^{0}$ (solid circles) and detector acceptance (dashed lines): (a) as a function of $-t$ and (b) as a function of $-u$.}
\label{fig:accept_t}
\end{figure}

Figure~\ref{fig:accept_t} shows the yields of $K^+\Sigma^{0}$ events as a function of $-t$ and $-u$ within the range of 1.169\,GeV/$c^2$ $< M_{\Lambda\gamma}<$ 1.217\,GeV/$c^2$. 
The acceptances are shown in the same figure as dashed lines. They were obtained by passing a sample of generated events through a GEANT3~\cite{Geant} model of the detector and applying the same selection criteria as used in the analysis.

\section{Photon Beam Asymmetry}

The event yields for the orthogonal orientations $Y_\parallel$ and $Y_\perp$ are given by Eqs.~\ref{eqn:y_para} and~\ref{eqn:y_perp}, where $\phi$ is the angle between a plane parallel to the laboratory floor and the $K^+$ production plane,  $\sigma_\text{0}$ is the unpolarized cross section, $A(\phi)$ is a function representing the detector acceptance, 
$N_\parallel$ ($N_\perp$) is the flux of photons,  $P_\parallel$ ($P_\perp$) is the magnitude of the photon beam polarization and $\Sigma$ is the beam asymmetry.

\begin{equation}
Y_{\parallel}(\phi)\propto N_{\parallel}[\sigma_\text{0}A(\phi)(1-P_{\parallel}\Sigma\,\text{cos}\,2\phi)]
\label{eqn:y_para}
\end{equation}
\begin{equation}
Y_{\perp}(\phi)\propto N_{\perp}[\sigma_\text{0}A(\phi)(1+P_{\perp}\Sigma\,\text{cos}\,2\phi)]
\label{eqn:y_perp}
\end{equation}

\begin{figure}[!ht]
\centering
\includegraphics[width=0.5\textwidth]{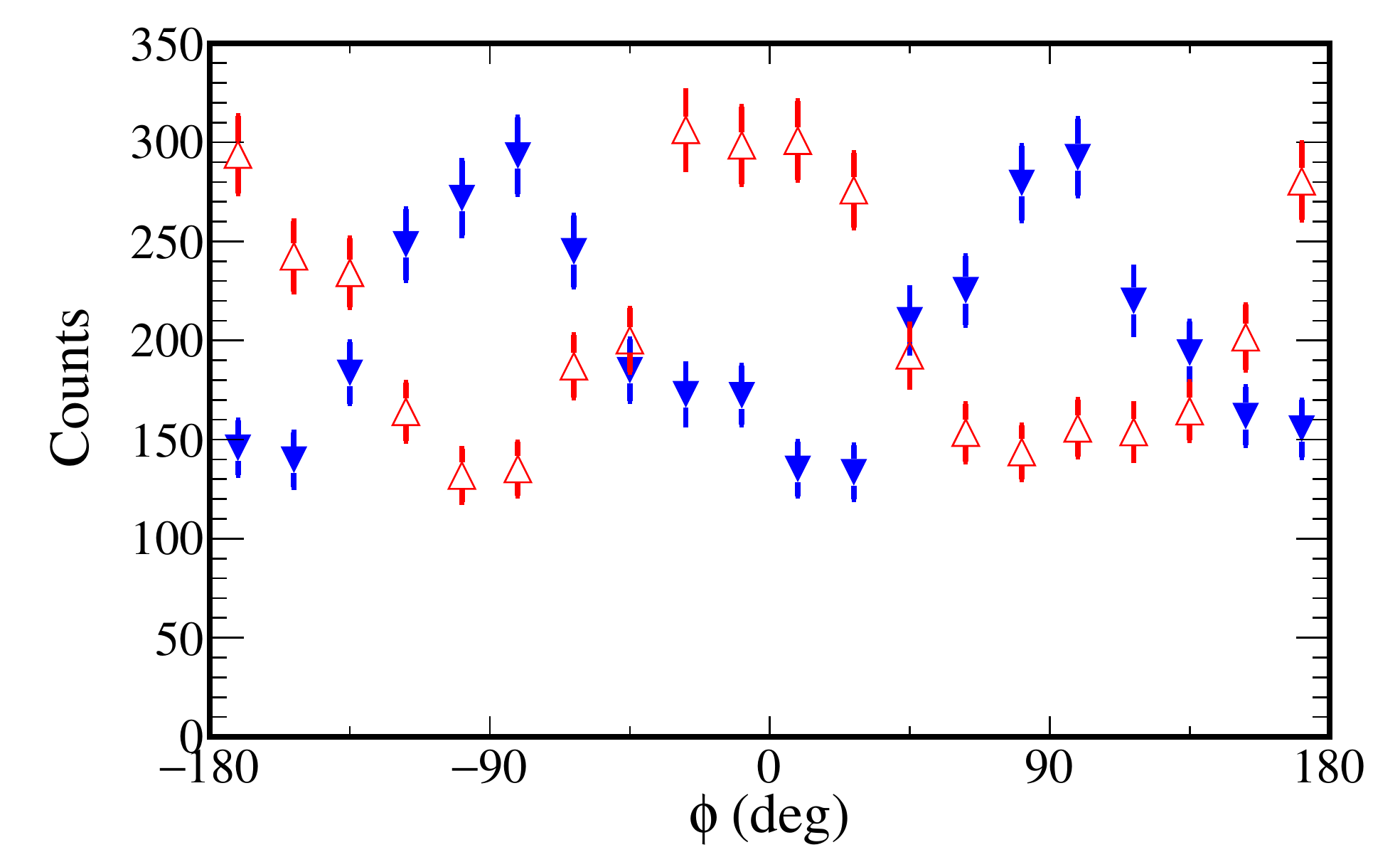}
\caption{ Yield of $\vec{\gamma} p\to K^{+}\Sigma^{0}$ events versus $\phi$ integrated over $t$ for the $90^{\circ}$ (open upward triangles) and $0^{\circ}$ (closed downward triangles) polarization orientations. }
\label{fig:t_0_PARA_PERP}
\end{figure}

\begin{figure}[!ht]
\centering
\includegraphics[width=0.5\textwidth]{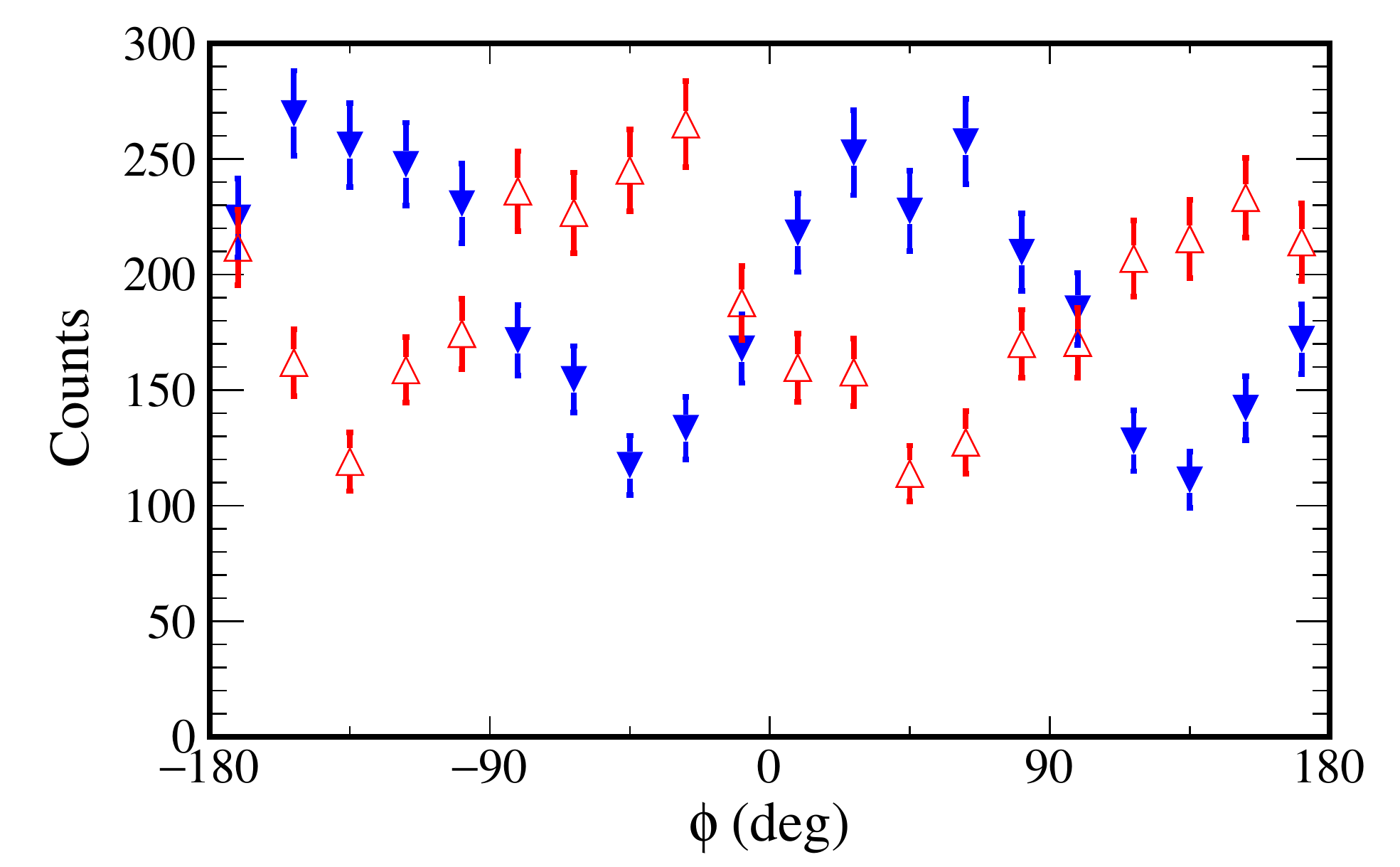}
\caption{ Yield of $\vec{\gamma} p\to K^{+}\Sigma^{0}$ events versus $\phi$ integrated over $t$ for the $45^{\circ}$ (open upward triangles) and $-45^{\circ}$ (closed downward triangles) polarization orientations. }
\label{fig:t_45_PARA_PERP}
\end{figure}

\begin{figure}[!ht]
\centering
\includegraphics[width=0.5\textwidth]{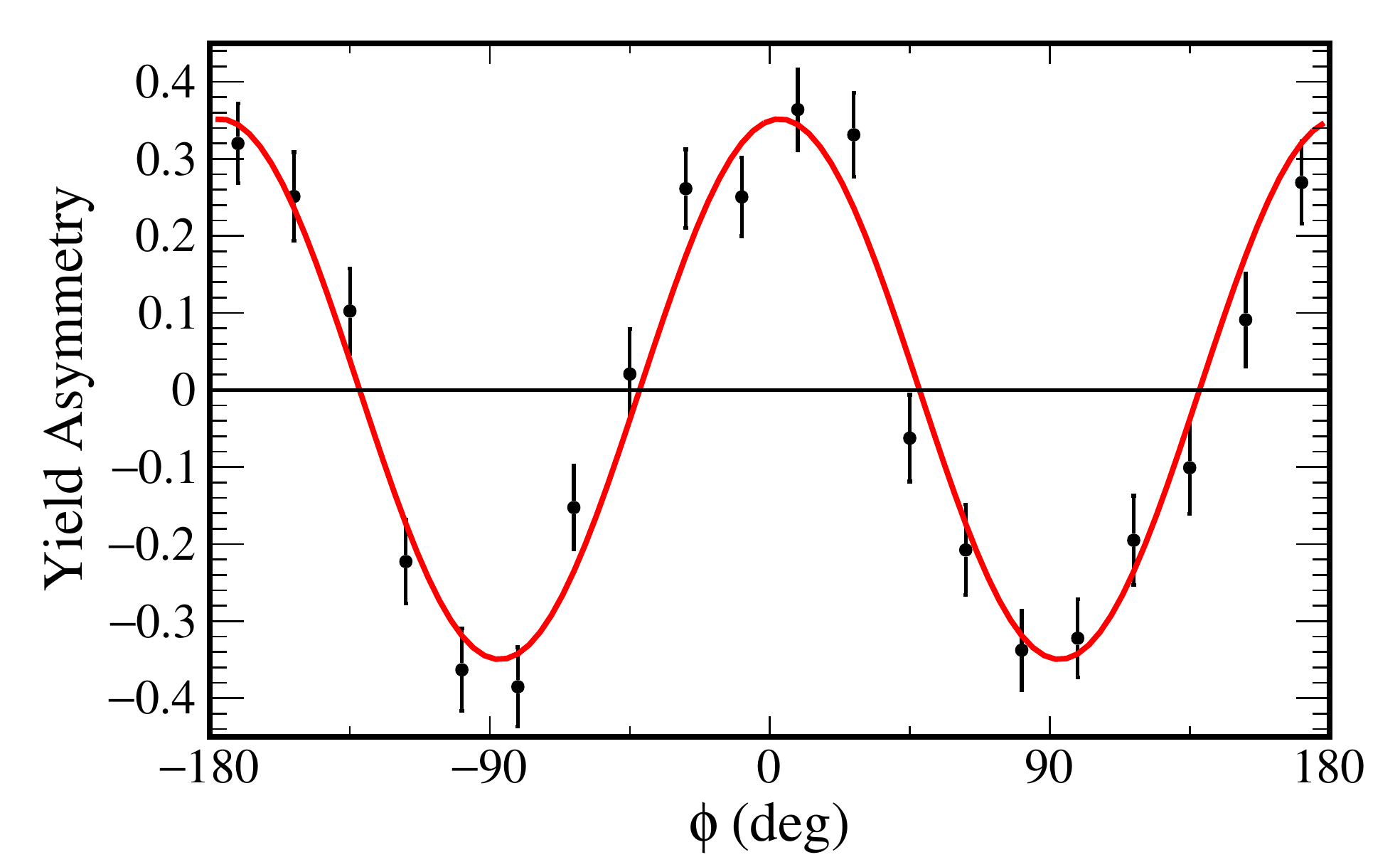}
\caption{The yield asymmetry for the 0/90 orientation set, corresponding to the data in Fig.~\ref{fig:t_0_PARA_PERP} with a fit of Eq.~\ref{eqn:asymm} (solid curve). See text for details. }
\label{fig:t_0_YA}
\end{figure}

\begin{figure}[!ht]
\centering
\includegraphics[width=0.5\textwidth]{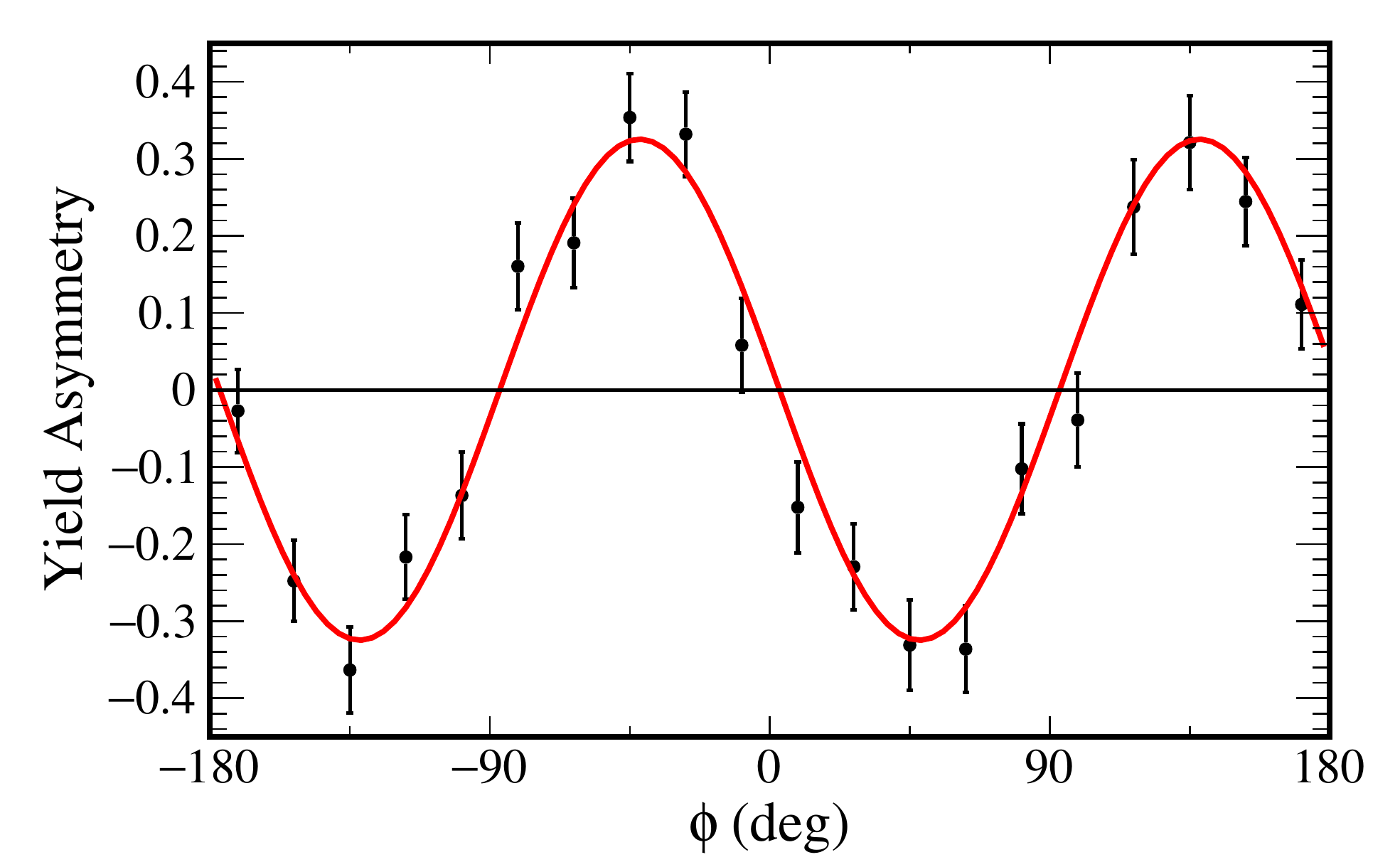}
\caption{The yield asymmetry for the -45/45 orientation set, corresponding to the data in Fig.~\ref{fig:t_45_PARA_PERP} with a fit of Eq.~\ref{eqn:asymm} (solid curve). See text for details. }
\label{fig:t_45_YA}
\end{figure}

Figure~\ref{fig:t_0_PARA_PERP} shows the yields for the photon polarization planes oriented at $0^{\circ}(Y_{\parallel})$ and $90^{\circ}(Y_{\perp})$, integrated over the $t$ region used in the analysis and Fig.~\ref{fig:t_45_PARA_PERP} shows the yields for the other orientation set, $-45^{\circ}(Y_{\parallel})$ and $45^{\circ}(Y_{\perp})$. Assuming that there is no background, these yields can be used to obtain a polarization-dependent yield asymmetry, given by\\
\begin{equation}
\label{eqn:asymm}
\frac{Y_{\perp}-F_{R}Y_{\parallel}}{Y_{\perp}+F_{R}Y_{\parallel}}=\frac{(P_{\perp}+P_{\parallel})\Sigma\,\text{cos}\,2(\phi-\phi_{0})}{2+(P_{\perp}-P_{\parallel})\Sigma \,\text{cos}\,2(\phi-\phi_{0})} 
\end{equation}
where $F_{R}=\frac{N_{\perp}}{N_{\parallel}}$ is the ratio of the integrated photon flux for the two orthogonal orientations.  A phase offset $\phi_{0}$ accounts for a possible small misalignment of the beam polarization from its nominal orientation and the additional $45^{\circ}$ offset for the -45/45 dataset.
The flux normalization ratio $F_{R}$ was found to be 1.038 for the 0/90 dataset and 0.995 for the -45/45 dataset. 
The yield asymmetry allows the beam asymmetry $\Sigma$ to be extracted without requiring any correction for instrumental acceptance.
The yield asymmetries for the 0/90 and -45/45 orientation sets are shown in Figs.~\ref{fig:t_0_YA} and~\ref{fig:t_45_YA} respectively.

After fitting the yield asymmetry with the function given in Eq.~\ref{eqn:asymm}, the beam asymmetry $\Sigma$ was extracted as the only free parameter in the fit. The yield asymmetry was measured in four bins of $t$, with roughly equal statistics in each bin.
The beam asymmetry values for the 0/90 and -45/45 orientations were combined using weighted averages.

Systematic uncertainties were estimated by varying the event selection criteria, the phase offset $\phi_{0}$, the flux normalization, and the minimum shower energy. 
They are listed in Tables~\ref{Tab:summarysyst_t} and \ref{Tab:summarysyst_u}.

For the event selection, the invariant mass cuts for $\pi^{-}p$ and $\Lambda\gamma$ were varied within the Gaussian 2$\sigma$ and 4$\sigma$ range, where $3\sigma$ is the nominal range. For the other cuts in the event selection, they are varied between ranges such that the signal yield was not allowed to change by more than 10\% from the nominal range to avoid statistical effects.
The systematic uncertainty due to the phase offset $\phi_{0}$ was found by letting $\phi_{0}$ be a free parameter in the fit and extracting beam asymmetry $\Sigma$ values.
The flux normalization was varied $\pm$5\% from the nominal value and the systematic uncertainty found using the corresponding $\Sigma$ values.

The minimum detection threshold for shower energy in the barrel calorimeter is 50\,MeV~\cite{Beattie:2018xsk}. The acceptance for radiated photons from low momentum $\Sigma^{0}$ decay is sensitive to this energy threshold at low $-t$. The systematic uncertainty was found by varying this minimum radiative photon energy to 55\,MeV and 60\,MeV. For the low $-u$ domain, the $\Sigma^{0}$ has high momentum leading to higher radiative photon energies, making the acceptance insensitive to the minimum shower energy around 50\,MeV. Therefore the systematic uncertainty due to this is estimated for the low $-t$ domain only.

The uncertainty from the 2\% background was estimated by measuring beam asymmetry for events in the region 1.23\,GeV/$c^{2} < M_{\Lambda\gamma} < 1.4$\,GeV/$c^{2}$. These are events from K$^+\Lambda$ combined with an uncorrelated shower. The systematic uncertainty from this background is 0.4\% for both $t$ and $u$ regions.

Since this reaction is studied in the fully exclusive final state, there is a potential bias arising from the non-uniform acceptance of decay products of the polarized $\Lambda$.  This leads the measured $\phi$ yields to be sensitive to unmeasured polarization observables of the recoiling hyperon~\cite{Paterson:2016vmc}.  A conservative estimate of the uncertainty due to this effect was made by convoluting the acceptance of the decay proton, obtained from detailed Monte Carlo simulations, with a range of polarization observables  spanning a conservative range of values.  The contribution of the hyperon decay dependence to the yield asymmetry was found to be 3\% or less for each bin in $t$.  A uniform 3\% systematic uncertainty was applied to all bins. The same approach was used for the $u$-channel production, for which 1.5\% uncertainty was obtained.

The dominant systematic uncertainties are due to the variation in event selection criteria.  A 2.1\% relative uncertainty in the measurement of the photon beam polarization comes from the combination of the 1.5\% systematic uncertainty in the instrument combined with the statistical uncertainty in the number of detected triplet events.  This uncertainty applies to the overall scale of the measured beam asymmetries and is not combined with the other uncertainties.

Table~\ref{Tab:summary_t} gives the average values of the beam asymmetry, together with the statistical and systematic uncertainties for the low $-t$ region. The combined systematic uncertainty for each bin in $t$ or $u$ is taken to be the larger of the systematic uncertainties from the two data sets, and the total uncertainties are found by adding the statistical and systematic errors in quadrature. 
 
\begin{figure}[!ht]
\begin{center}
\includegraphics[width=0.5\textwidth]{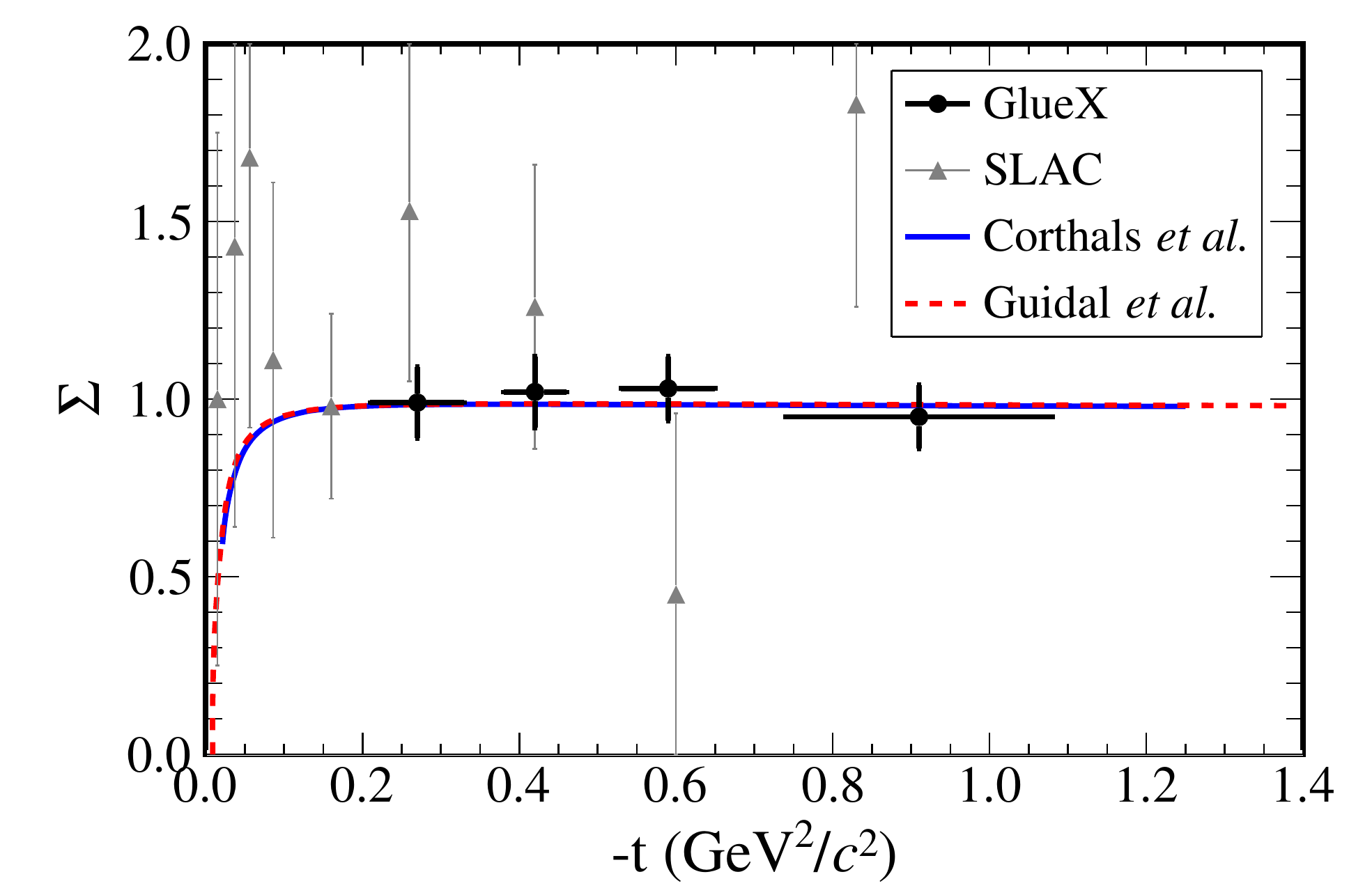}
\caption{ The beam asymmetry $\Sigma$ for $\vec{\gamma} p\to K^+\Sigma^0$ as a function of $-t$. The results from the 0/90 and -45/45 data sets are averaged (solid circles) where horizontal error bars indicate the RMS widths of the $t$ bins and vertical error bars represent statistical and systematic uncertainties added in quadrature.  An additional 2.1\% overall relative polarization uncertainty is not included.  The triangles are previous SLAC results~\cite{Quinn} at $E_{\gamma} = 16$\,GeV, the 
curves show predictions from RPR-2007~\cite{Ghent,Ghent2} (solid) and Guidal \textit{et al.}~\cite{Guidal} (dashed) at $E_{\gamma} = 8.5$\,GeV.}
\label{fig:BA_t}
\end{center}
\end{figure}

The extracted beam asymmetry results shown in Fig.~\ref{fig:BA_t} are close to unity within errors in all four $t$ bins. The mean value of $\Sigma$ over the entire measured $t$ range is found to be $\Sigma=1.00\pm 0.04\text{(stat)}\pm 0.03\text{(syst)}\pm 0.02\text{(pol)}$. From Eq.~\ref{eqn:parity} it follows that natural-parity exchange dominates in the photoproduction of $K^+\Sigma^0$. This result is consistent with the theoretical predictions from RPR-2007~\cite{Ghent,Ghent2} and Guidal \textit{et al.}~\cite{Guidal} where $K^+\Sigma^0$ photoproduction proceeds via exchange of $K^{*}(892)$, the lowest member of the linear Regge trajectory for natural-parity exchange.

The beam asymmetry for the measured low $-u$ region, $-u < 2.0$ (GeV/$c$)$^{2}$, is found to be 
$\Sigma = 0.41~\pm~0.07\text{(stat)}~\pm~0.06\text{(syst)} ~\pm~0.02\text{(pol)}$
at an average value of $-u~=~0.53~\pm~0.34\,(\text{GeV}/c)^{2}$.

\begin{table} [ht]
\caption{Summary of systematic uncertainties for the low $-t\, (0.1 < -t < 1.4$ (GeV/$c$)$^{2}$) region.}
\begin{center}
\begin{tabular}{@{} l  @{\hspace*{5mm}} c @{\hspace*{5mm}} c @{\hspace*{5mm}} c}
\hline
\hline
\multicolumn{1}{c}{} &\multicolumn{1}{c}{} & \multicolumn{1}{c}{}\\
Source & 0/90 Set & -45/45 Set \\ 
\hline
Event selection & 3.1-5.9\% & 3.0-5.3\%\\
Phase offset & 0.1\% & 0.7\% \\
Flux normalization  & 0.5\% & 0.4\% \\
Minimum shower energy         & 2.6\% & 2.9\% \\
Background & 0.4\% & 0.4\% \\
Non-uniform acceptance & 3.0\% & 3.0\% \\
\hline
Total & 5.1-7.1\% & 5.2-6.8\% \\
\hline
\hline
\end{tabular}
\label{Tab:summarysyst_t}
\end{center}
\end{table}

\begin{table} [ht]
\caption{Summary of systematic uncertainties for the low $-u\,(-u < 2.0$ (GeV/$c$)$^{2})$ region.}
\begin{center}
\begin{tabular}{@{} l  @{\hspace*{5mm}} c @{\hspace*{5mm}} c @{\hspace*{5mm}} c}
\hline
\hline
\multicolumn{1}{c}{} &\multicolumn{1}{c}{} & \multicolumn{1}{c}{}\\
Source & 0/90 Set & -45/45 Set \\ 
\hline
Event selection & 5.0\% & 4.0\%\\
Phase offset & 2.2\% & 2.1\% \\
Flux normalization & 0.6\% & 0.2\% \\
Background & 0.4\% & 0.4\% \\
Non-uniform acceptance & 1.5\% & 1.5\% \\
\hline
Total & 5.7\% & 4.8\% \\
\hline
\hline
\end{tabular}
\label{Tab:summarysyst_u}
\end{center}
\end{table}

\begin{table} [ht]
\caption{ Average beam asymmetry $\Sigma$ for the low $-t$ region with statistical and systematic uncertainties.}
\begin{center}
\begin{tabular}{@{} c  @{\hspace*{15mm}} c}
\hline
\hline
$-t$  ((GeV/$c$)$^{2}$) & $\Sigma$\\
\hline
0.27 & 0.99~$\pm$~0.08~$\pm$~0.07\\
0.42 & 1.02~$\pm$~0.07~$\pm$~0.07\\
0.59 & 1.03~$\pm$~0.07~$\pm$~0.05\\
0.91 & 0.95~$\pm$~0.07~$\pm$~0.05\\
\hline
\hline
\end{tabular}
\label{Tab:summary_t}
\end{center}
\end{table}

\section{Conclusions}

We present experimental results for the first measurement of the photon beam asymmetry $\Sigma$ in the exclusive reaction  $\vec{\gamma} p \to K^{+}\Sigma^{0}$ beyond the baryon resonance region, which have significantly higher precision than the earlier SLAC measurement~\cite{Quinn}. The measured beam asymmetry as a function of $t$ is consistent, within a few percent, with unity and with model predictions from Refs.~\cite{Guidal, Ghent, Ghent2}, suggesting a dominant natural parity exchange. The beam asymmetry for the region of $-u < 2.0~(\mathrm{GeV}/c)^2$ has never been extracted before. An average beam asymmetry of 0.41~$\pm$~0.09 for the $u$ interval is obtained. 
In this kinematic domain, $u$-channel hyperon exchanges of both $\Sigma$ ($J=1/2$ trajectory) and $Y^{*}$ ($J=3/2$ trajectory) contribute to the production of the $K^{+}\Sigma^{0}$ final state. Currently there is no prediction for the beam asymmetry as a function of $u$. These results place significant new constraints on photoproduction models for strangeness-exchange reactions.

\vspace*{1cm}
\section*{Acknowledgements}
We would like to acknowledge the outstanding efforts of the staff of the Accelerator and the Physics Division at Jefferson Lab that made the experiment possible. We appreciate useful communication with Jan Ryckebusch, Michel Guidal and Gary Goldstein.

This work was supported in part by the U.S. Department of Energy, the U.S. National Science Foundation, the German Research Foundation, GSI Helmholtzzentrum f\"ur Schwerionenforschung GmbH, the Natural Sciences and Engineering Research Council of Canada, the Russian Foundation for Basic Research, the UK Science and Technology Facilities Council, the Chilean Comisi\'{o}n Nacional de Investigaci\'{o}n Cient\'{i}fica y Tecnol\'{o}gica, the National Natural Science Foundation of China and the China Scholarship Council. This material is based upon work supported by the U.S. Department of Energy, Office of Science, Office of Nuclear Physics under contract DE-AC05-06OR23177.

\bibliography{kpsig}

\begin{thebibliography}{22}%
\makeatletter
\providecommand \@ifxundefined [1]{%
 \@ifx{#1\undefined}
}%
\providecommand \@ifnum [1]{%
 \ifnum #1\expandafter \@firstoftwo
 \else \expandafter \@secondoftwo
 \fi
}%
\providecommand \@ifx [1]{%
 \ifx #1\expandafter \@firstoftwo
 \else \expandafter \@secondoftwo
 \fi
}%
\providecommand \natexlab [1]{#1}%
\providecommand \enquote  [1]{``#1''}%
\providecommand \bibnamefont  [1]{#1}%
\providecommand \bibfnamefont [1]{#1}%
\providecommand \citenamefont [1]{#1}%
\providecommand \href@noop [0]{\@secondoftwo}%
\providecommand \href [0]{\begingroup \@sanitize@url \@href}%
\providecommand \@href[1]{\@@startlink{#1}\@@href}%
\providecommand \@@href[1]{\endgroup#1\@@endlink}%
\providecommand \@sanitize@url [0]{\catcode `\\12\catcode `\$12\catcode
  `\&12\catcode `\#12\catcode `\^12\catcode `\_12\catcode `\%12\relax}%
\providecommand \@@startlink[1]{}%
\providecommand \@@endlink[0]{}%
\providecommand \url  [0]{\begingroup\@sanitize@url \@url }%
\providecommand \@url [1]{\endgroup\@href {#1}{\urlprefix }}%
\providecommand \urlprefix  [0]{URL }%
\providecommand \Eprint [0]{\href }%
\providecommand \doibase [0]{http://dx.doi.org/}%
\providecommand \selectlanguage [0]{\@gobble}%
\providecommand \bibinfo  [0]{\@secondoftwo}%
\providecommand \bibfield  [0]{\@secondoftwo}%
\providecommand \translation [1]{[#1]}%
\providecommand \BibitemOpen [0]{}%
\providecommand \bibitemStop [0]{}%
\providecommand \bibitemNoStop [0]{.\EOS\space}%
\providecommand \EOS [0]{\spacefactor3000\relax}%
\providecommand \BibitemShut  [1]{\csname bibitem#1\endcsname}%
\let\auto@bib@innerbib\@empty
\bibitem [{\citenamefont {Quinn}\ \emph {et~al.}(1975)\citenamefont {Quinn},
  \citenamefont {Rutherfoord}, \citenamefont {Shupe}, \citenamefont {Sherden},
  \citenamefont {Siemann},\ and\ \citenamefont {Sinclair}}]{Quinn1975}%
  \BibitemOpen
  \bibfield  {author} {\bibinfo {author} {\bibfnamefont {D.~J.}\ \bibnamefont
  {Quinn}}, \bibinfo {author} {\bibfnamefont {J.~P.}\ \bibnamefont
  {Rutherfoord}}, \bibinfo {author} {\bibfnamefont {M.~A.}\ \bibnamefont
  {Shupe}}, \bibinfo {author} {\bibfnamefont {D.~J.}\ \bibnamefont {Sherden}},
  \bibinfo {author} {\bibfnamefont {R.~H.}\ \bibnamefont {Siemann}}, \ and\
  \bibinfo {author} {\bibfnamefont {C.~K.}\ \bibnamefont {Sinclair}},\ }\href
  {\doibase 10.1103/PhysRevLett.34.543} {\bibfield  {journal} {\bibinfo
  {journal} {Phys. Rev. Lett.}\ }\textbf {\bibinfo {volume} {34}},\ \bibinfo
  {pages} {543} (\bibinfo {year} {1975})}\BibitemShut {NoStop}%
\bibitem [{\citenamefont {Quinn}\ \emph {et~al.}(1979)\citenamefont {Quinn},
  \citenamefont {Rutherfoord}, \citenamefont {Shupe}, \citenamefont {Sherden},
  \citenamefont {Siemann},\ and\ \citenamefont {Sinclair}}]{Quinn}%
  \BibitemOpen
  \bibfield  {author} {\bibinfo {author} {\bibfnamefont {D.~J.}\ \bibnamefont
  {Quinn}}, \bibinfo {author} {\bibfnamefont {J.~P.}\ \bibnamefont
  {Rutherfoord}}, \bibinfo {author} {\bibfnamefont {M.~A.}\ \bibnamefont
  {Shupe}}, \bibinfo {author} {\bibfnamefont {D.~J.}\ \bibnamefont {Sherden}},
  \bibinfo {author} {\bibfnamefont {R.~H.}\ \bibnamefont {Siemann}}, \ and\
  \bibinfo {author} {\bibfnamefont {C.~K.}\ \bibnamefont {Sinclair}},\ }\href
  {\doibase 10.1103/PhysRevD.20.1553} {\bibfield  {journal} {\bibinfo
  {journal} {Phys. Rev.}\ }\textbf {\bibinfo {volume} {D20}},\ \bibinfo {pages}
  {1553} (\bibinfo {year} {1979})}\BibitemShut {NoStop}%
\bibitem [{\citenamefont {Goldstein}\ \emph {et~al.}(1973)\citenamefont
  {Goldstein}, \citenamefont {Owens~III},\ and\ \citenamefont
  {Rutherfoord}}]{Gold}%
  \BibitemOpen
  \bibfield  {author} {\bibinfo {author} {\bibfnamefont {G.~R.}\ \bibnamefont
  {Goldstein}}, \bibinfo {author} {\bibfnamefont {J.~F.}\ \bibnamefont
  {Owens~III}}, \ and\ \bibinfo {author} {\bibfnamefont {J.}~\bibnamefont
  {Rutherfoord}},\ }\href {\doibase 10.1016/0550-3213(73)90312-X} {\bibfield
  {journal} {\bibinfo  {journal} {Nucl. Phys.}\ }\textbf {\bibinfo {volume}
  {B53}},\ \bibinfo {pages} {197} (\bibinfo {year} {1973})}\BibitemShut
  {NoStop}%
\bibitem [{\citenamefont {Levy}\ \emph {et~al.}(1973)\citenamefont {Levy},
  \citenamefont {Majerotto},\ and\ \citenamefont {Read}}]{Levy}%
  \BibitemOpen
  \bibfield  {author} {\bibinfo {author} {\bibfnamefont {N.}~\bibnamefont
  {Levy}}, \bibinfo {author} {\bibfnamefont {W.}~\bibnamefont {Majerotto}}, \
  and\ \bibinfo {author} {\bibfnamefont {B.~J.}\ \bibnamefont {Read}},\ }\href
  {\doibase 10.1016/0550-3213(73)90393-3} {\bibfield  {journal} {\bibinfo
  {journal} {Nucl. Phys.}\ }\textbf {\bibinfo {volume} {B55}},\ \bibinfo
  {pages} {493} (\bibinfo {year} {1973})}\BibitemShut {NoStop}%
\bibitem [{\citenamefont {Guidal}\ \emph {et~al.}(1997)\citenamefont {Guidal},
  \citenamefont {Laget},\ and\ \citenamefont {Vanderhaeghen}}]{Guidal}%
  \BibitemOpen
  \bibfield  {author} {\bibinfo {author} {\bibfnamefont {M.}~\bibnamefont
  {Guidal}}, \bibinfo {author} {\bibfnamefont {J.~M.}\ \bibnamefont {Laget}}, \
  and\ \bibinfo {author} {\bibfnamefont {M.}~\bibnamefont {Vanderhaeghen}},\
  }\href {\doibase 10.1016/S0375-9474(97)00612-X} {\bibfield  {journal}
  {\bibinfo  {journal} {Nucl. Phys.}\ }\textbf {\bibinfo {volume} {A627}},\
  \bibinfo {pages} {645} (\bibinfo {year} {1997})}\BibitemShut {NoStop}%
\bibitem [{\citenamefont {Corthals}\ \emph {et~al.}(2006)\citenamefont
  {Corthals}, \citenamefont {Ryckebusch},\ and\ \citenamefont
  {Van~Cauteren}}]{Ghent}%
  \BibitemOpen
  \bibfield  {author} {\bibinfo {author} {\bibfnamefont {T.}~\bibnamefont
  {Corthals}}, \bibinfo {author} {\bibfnamefont {J.}~\bibnamefont
  {Ryckebusch}}, \ and\ \bibinfo {author} {\bibfnamefont {T.}~\bibnamefont
  {Van~Cauteren}},\ }\href {\doibase 10.1103/PhysRevC.73.045207} {\bibfield
  {journal} {\bibinfo  {journal} {Phys. Rev.}\ }\textbf {\bibinfo {volume}
  {C73}},\ \bibinfo {pages} {045207} (\bibinfo {year} {2006})}\BibitemShut
  {NoStop}%
\bibitem [{\citenamefont {Corthals}\ \emph {et~al.}(2007)\citenamefont
  {Corthals}, \citenamefont {Ireland}, \citenamefont {Van~Cauteren},\ and\
  \citenamefont {Ryckebusch}}]{Ghent2}%
  \BibitemOpen
  \bibfield  {author} {\bibinfo {author} {\bibfnamefont {T.}~\bibnamefont
  {Corthals}}, \bibinfo {author} {\bibfnamefont {D.~G.}\ \bibnamefont
  {Ireland}}, \bibinfo {author} {\bibfnamefont {T.}~\bibnamefont
  {Van~Cauteren}}, \ and\ \bibinfo {author} {\bibfnamefont {J.}~\bibnamefont
  {Ryckebusch}},\ }\href {\doibase 10.1103/PhysRevC.75.045204} {\bibfield
  {journal} {\bibinfo  {journal} {Phys. Rev.}\ }\textbf {\bibinfo {volume}
  {C75}},\ \bibinfo {pages} {045204} (\bibinfo {year} {2007})}\BibitemShut
  {NoStop}%
\bibitem [{\citenamefont {Zegers}\ \emph {et~al.}(2003)\citenamefont {Zegers}
  \emph {et~al.}}]{Zegers:2003ux}%
  \BibitemOpen
  \bibfield  {author} {\bibinfo {author} {\bibfnamefont {R.~G.~T.}\
  \bibnamefont {Zegers}} \emph {et~al.} (\bibinfo {collaboration} {LEPS}),\
  }\href {\doibase 10.1103/PhysRevLett.91.092001} {\bibfield  {journal}
  {\bibinfo  {journal} {Phys. Rev. Lett.}\ }\textbf {\bibinfo {volume} {91}},\
  \bibinfo {pages} {092001} (\bibinfo {year} {2003})}\BibitemShut {NoStop}%
\bibitem [{\citenamefont {Kohri}\ \emph {et~al.}(2006)\citenamefont {Kohri}
  \emph {et~al.}}]{Kohri:2006yx}%
  \BibitemOpen
  \bibfield  {author} {\bibinfo {author} {\bibfnamefont {H.}~\bibnamefont
  {Kohri}} \emph {et~al.} (\bibinfo {collaboration} {LEPS}),\ }\href {\doibase
  10.1103/PhysRevLett.97.082003} {\bibfield  {journal} {\bibinfo  {journal}
  {Phys. Rev. Lett.}\ }\textbf {\bibinfo {volume} {97}},\ \bibinfo {pages}
  {082003} (\bibinfo {year} {2006})}\BibitemShut {NoStop}%
\bibitem [{\citenamefont {Lleres}\ \emph {et~al.}(2007)\citenamefont {Lleres}
  \emph {et~al.}}]{Lleres:2007tx}%
  \BibitemOpen
  \bibfield  {author} {\bibinfo {author} {\bibfnamefont {A.}~\bibnamefont
  {Lleres}} \emph {et~al.},\ }\href {\doibase 10.1140/epja/i2006-10167-8}
  {\bibfield  {journal} {\bibinfo  {journal} {Eur. Phys. J.}\ }\textbf
  {\bibinfo {volume} {A31}},\ \bibinfo {pages} {79} (\bibinfo {year}
  {2007})}\BibitemShut {NoStop}%
\bibitem [{\citenamefont {Lleres}\ \emph {et~al.}(2009)\citenamefont {Lleres}
  \emph {et~al.}}]{Lleres:2008em}%
  \BibitemOpen
  \bibfield  {author} {\bibinfo {author} {\bibfnamefont {A.}~\bibnamefont
  {Lleres}} \emph {et~al.} (\bibinfo {collaboration} {GRAAL}),\ }\href
  {\doibase 10.1140/epja/i2008-10713-4} {\bibfield  {journal} {\bibinfo
  {journal} {Eur. Phys. J.}\ }\textbf {\bibinfo {volume} {A39}},\ \bibinfo
  {pages} {149} (\bibinfo {year} {2009})}\BibitemShut {NoStop}%
\bibitem [{\citenamefont {Paterson}\ \emph {et~al.}(2016)\citenamefont
  {Paterson} \emph {et~al.}}]{Paterson:2016vmc}%
  \BibitemOpen
  \bibfield  {author} {\bibinfo {author} {\bibfnamefont {C.~A.}\ \bibnamefont
  {Paterson}} \emph {et~al.} (\bibinfo {collaboration} {CLAS}),\ }\href
  {\doibase 10.1103/PhysRevC.93.065201} {\bibfield  {journal} {\bibinfo
  {journal} {Phys. Rev.}\ }\textbf {\bibinfo {volume} {C93}},\ \bibinfo {pages}
  {065201} (\bibinfo {year} {2016})}\BibitemShut {NoStop}%
\bibitem [{\citenamefont {Dugger}\ \emph {et~al.}(2017)\citenamefont {Dugger}
  \emph {et~al.}}]{Dugger:2017zoq}%
  \BibitemOpen
  \bibfield  {author} {\bibinfo {author} {\bibfnamefont {M.}~\bibnamefont
  {Dugger}} \emph {et~al.},\ }\href {\doibase 10.1016/j.nima.2017.05.026}
  {\bibfield  {journal} {\bibinfo  {journal} {Nucl. Instrum. Meth.}\ }\textbf
  {\bibinfo {volume} {A867}},\ \bibinfo {pages} {115} (\bibinfo {year}
  {2017})}\BibitemShut {NoStop}%
\bibitem [{\citenamefont {Barbosa}\ \emph {et~al.}(2015)\citenamefont
  {Barbosa}, \citenamefont {Hutton}, \citenamefont {Sitnikov}, \citenamefont
  {Somov}, \citenamefont {Somov},\ and\ \citenamefont
  {Tolstukhin}}]{Barbosa:2015bga}%
  \BibitemOpen
  \bibfield  {author} {\bibinfo {author} {\bibfnamefont {F.}~\bibnamefont
  {Barbosa}}, \bibinfo {author} {\bibfnamefont {C.}~\bibnamefont {Hutton}},
  \bibinfo {author} {\bibfnamefont {A.}~\bibnamefont {Sitnikov}}, \bibinfo
  {author} {\bibfnamefont {A.}~\bibnamefont {Somov}}, \bibinfo {author}
  {\bibfnamefont {S.}~\bibnamefont {Somov}}, \ and\ \bibinfo {author}
  {\bibfnamefont {I.}~\bibnamefont {Tolstukhin}},\ }\href {\doibase
  10.1016/j.nima.2015.06.012} {\bibfield  {journal} {\bibinfo  {journal} {Nucl.
  Instrum. Meth.}\ }\textbf {\bibinfo {volume} {A795}},\ \bibinfo {pages} {376}
  (\bibinfo {year} {2015})}\BibitemShut {NoStop}%
\bibitem [{\citenamefont {Pooser}\ \emph {et~al.}(2019)\citenamefont {Pooser}
  \emph {et~al.}}]{Pooser:2019rhu}%
  \BibitemOpen
  \bibfield  {author} {\bibinfo {author} {\bibfnamefont {E.}~\bibnamefont
  {Pooser}} \emph {et~al.},\ }\href {\doibase 10.1016/j.nima.2019.02.029}
  {\bibfield  {journal} {\bibinfo  {journal} {Nucl. Instrum. Meth.}\ }\textbf
  {\bibinfo {volume} {A927}},\ \bibinfo {pages} {330} (\bibinfo {year}
  {2019})}\BibitemShut {NoStop}%
\bibitem [{\citenamefont {Jarvis}\ \emph {et~al.}(2020)\citenamefont {Jarvis}
  \emph {et~al.}}]{CDC_nim}%
  \BibitemOpen
  \bibfield  {author} {\bibinfo {author} {\bibfnamefont {N.}~\bibnamefont
  {Jarvis}} \emph {et~al.},\ }\href {\doibase 10.1016/j.nima.2020.163727}
  {\bibfield  {journal} {\bibinfo  {journal} {Nucl.\ Instrum.\ Meth.}\ }\textbf
  {\bibinfo {volume} {A962}},\ \bibinfo {pages} {163727} (\bibinfo {year}
  {2020})}\BibitemShut {NoStop}%
\bibitem [{\citenamefont {Beattie}\ \emph {et~al.}(2018)\citenamefont {Beattie}
  \emph {et~al.}}]{Beattie:2018xsk}%
  \BibitemOpen
  \bibfield  {author} {\bibinfo {author} {\bibfnamefont {T.~D.}\ \bibnamefont
  {Beattie}} \emph {et~al.},\ }\href {\doibase 10.1016/j.nima.2018.04.006}
  {\bibfield  {journal} {\bibinfo  {journal} {Nucl. Instrum. Meth.}\ }\textbf
  {\bibinfo {volume} {A896}},\ \bibinfo {pages} {24} (\bibinfo {year}
  {2018})}\BibitemShut {NoStop}%
\bibitem [{\citenamefont {Pentchev}\ \emph {et~al.}(2017)\citenamefont
  {Pentchev}, \citenamefont {Barbosa}, \citenamefont {Berdnikov}, \citenamefont
  {Butler}, \citenamefont {Furletov}, \citenamefont {Robison},\ and\
  \citenamefont {Zihlmann}}]{Pentchev:2017omk}%
  \BibitemOpen
  \bibfield  {author} {\bibinfo {author} {\bibfnamefont {L.}~\bibnamefont
  {Pentchev}}, \bibinfo {author} {\bibfnamefont {F.}~\bibnamefont {Barbosa}},
  \bibinfo {author} {\bibfnamefont {V.}~\bibnamefont {Berdnikov}}, \bibinfo
  {author} {\bibfnamefont {D.}~\bibnamefont {Butler}}, \bibinfo {author}
  {\bibfnamefont {S.}~\bibnamefont {Furletov}}, \bibinfo {author}
  {\bibfnamefont {L.}~\bibnamefont {Robison}}, \ and\ \bibinfo {author}
  {\bibfnamefont {B.}~\bibnamefont {Zihlmann}},\ }\href {\doibase
  10.1016/j.nima.2016.04.076} {\bibfield  {journal} {\bibinfo  {journal} {Nucl.
  Instrum. Meth.}\ }\textbf {\bibinfo {volume} {A845}},\ \bibinfo {pages} {281}
  (\bibinfo {year} {2017})}\BibitemShut {NoStop}%
\bibitem [{\citenamefont {Moriya}\ \emph {et~al.}(2013)\citenamefont {Moriya}
  \emph {et~al.}}]{Moriya:2013aja}%
  \BibitemOpen
  \bibfield  {author} {\bibinfo {author} {\bibfnamefont {K.}~\bibnamefont
  {Moriya}} \emph {et~al.},\ }\href {\doibase 10.1016/j.nima.2013.05.109}
  {\bibfield  {journal} {\bibinfo  {journal} {Nucl. Instrum. Meth.}\ }\textbf
  {\bibinfo {volume} {A726}},\ \bibinfo {pages} {60} (\bibinfo {year}
  {2013})}\BibitemShut {NoStop}%
\bibitem [{\citenamefont {Barsotti}\ and\ \citenamefont
  {Shepherd}()}]{Barsotti:2020}%
  \BibitemOpen
  \bibfield  {author} {\bibinfo {author} {\bibfnamefont {R.}~\bibnamefont
  {Barsotti}}\ and\ \bibinfo {author} {\bibfnamefont {M.~R.}\ \bibnamefont
  {Shepherd}},\ }\href@noop {} {\bibfield  {journal} {\bibinfo  {journal}
  {submitted to JINST,}\ }}\Eprint {http://arxiv.org/abs/2002.09530}
  {arXiv:2002.09530 [physics.data-an]} \BibitemShut {NoStop}%
\bibitem [{\citenamefont {Adhikari}\ \emph {et~al.}(2019)\citenamefont
  {Adhikari} \emph {et~al.}}]{Adhikari:2019gfa}%
  \BibitemOpen
  \bibfield  {author} {\bibinfo {author} {\bibfnamefont {S.}~\bibnamefont
  {Adhikari}} \emph {et~al.} (\bibinfo {collaboration} {GlueX}),\ }\href
  {\doibase 10.1103/PhysRevC.100.052201} {\bibfield  {journal} {\bibinfo
  {journal} {Phys. Rev.}\ }\textbf {\bibinfo {volume} {C100}},\ \bibinfo
  {pages} {052201} (\bibinfo {year} {2019})}\BibitemShut {NoStop}%
\bibitem [{\citenamefont {Brun}\ \emph {et~al.}(1978)\citenamefont {Brun} \emph
  {et~al.}}]{Geant}%
  \BibitemOpen
  \bibfield  {author} {\bibinfo {author} {\bibfnamefont {R.}~\bibnamefont
  {Brun}} \emph {et~al.},\ }\href@noop {} {\bibfield  {journal} {\bibinfo
  {journal} {Report No. CERN-DD-78-2-REV}\ } (\bibinfo {year}
  {1978})}\BibitemShut {NoStop}%
\end{thebibliography}%

\end{document}